\documentclass[aps,pre,twocolumn,superscriptaddress]{revtex4}

\usepackage{dcolumn}
\usepackage{amsmath}
\usepackage{amssymb}
\usepackage{commath}
\usepackage{xfrac}
\usepackage{braket}
\usepackage{upgreek,CJK}
\usepackage{mathtools}
\usepackage{graphicx}

\def\be{\begin{equation}}
\def\ee{\end{equation}}
\def\ba{\begin{eqnarray}}
\def\ea{\end{eqnarray}}

\begin{document}


\title{Generalized Wigner-von Neumann entropy and its typicality}


\begin{CJK}{UTF8}{gbsn}
\author{Zhigang Hu(胡志刚)}
\affiliation{International Center for Quantum Materials, School of Physics, Peking University, Beijing 100871, China}

\author{Zhenduo Wang(王朕铎)}
\affiliation{International Center for Quantum Materials, School of Physics, Peking University, Beijing 100871, China}
\author{Biao Wu(吴飙)}
\email{wubiao@pku.edu.cn}
\affiliation{International Center for Quantum Materials, School of Physics, Peking University, Beijing 100871, China}
\affiliation{Collaborative Innovation Center of Quantum Matter, Beijing 100871, China}
\affiliation{Wilczek Quantum Center, School of Physics and Astronomy, Shanghai Jiao Tong University, Shanghai 200240, China}


\date{\today}

\begin{abstract}
We propose  a generalization of the quantum entropy introduced
by Wigner and von Neumann [Zeitschrift f\"ur Physik 57, 30 (1929)]. Our generalization is applicable
to both quantum pure states and mixed states. When the dimension $N$ of the  Hilbert space  is large,
 this generalized Wigner-von Neumann (GWvN) entropy becomes independent
of choices of basis and is  asymptotically equal  to $\ln N$ in the sense  of typicality. The dynamic evolution of our entropy
is also typical, reminiscent of quantum H theorem proved by von Neumann.  For a composite system, 
the GWvN entropy is typically additive;  for the microcanonical ensemble,
it is equivalent to the Boltzmann entropy; for a system entangled with environment,
it is consistent with the familiar von Neumann entropy, which is zero for pure states. In addition, 
the GWvN entropy can be used to derive the Gibbs ensemble.
\end{abstract}


\maketitle
\end{CJK}
\section{Introduction}
As an empirical understanding of macroscopic irreversible processes, the second law of thermodynamics
has stood the test of time. However,  its microscopic origin is still perplexing physicists. Boltzmann gave us a
clear understanding how the second law could arise from the time-reversible Newton's laws of motion.
Unfortunately, we now know that macroscopic objects are made of microscopic particles that
obey the laws of quantum mechanics.  It is therefore desirable and imperative to understand
the second law in the perspective of quantum
dynamics
~\cite{Neumann1929,Neumann2010,Srednicki1994,Gemmer2003,Popescu2006,
Rigol2008Nature,reimann2008,Gogolin2010,
Goldstein2013,Goldstein2014,Han2015,Tian,Gogolin2016,Tian2018PRL,Tian2018PRE}.

In 1929, von Neumann made the first attempt  to understand the second law quantum mechanically
by proving quantum ergodic theorem and quantum H theorem~\cite{Neumann1929,Neumann2010,Goldstein2010}.
In  proving the H theorem, von Neumann introduced an entropy for quantum pure states but acknowledged that this definition came
from Wigner's unpublished work\cite{Neumann1929,Neumann2010}. Therefore, we call such a quantum entropy
Wigner-von Neumann (WvN) entropy. The well-known von Neumann entropy was not used because it is always
zero for quantum pure states and can not describe the relaxation and fluctuations in isolated macroscopic quantum systems.
After introducing WvN entropy, von Neumann showed with the help of typicality arguments that for overwhelming
majority of basis the H-theorem holds without exception for all states.
The typicality argument \cite{Lloyd2013,Goldstein2006,Reimann2007,Reimann2010,Tasaki2015} is
mathematically known as measure concentration \cite{1986a} and Levy's lemma \cite{Popescu2006}.

However, WvN entropy involves vaguely-defined coarse graining and it does not apply to systems with spins.
It was shown in Ref. \cite{Han2015} that  the quantum H theorem still holds for a generalized WvN entropy that
does not involve coarse graining.  In this work we generalize WvN entropy further so that it applies to
any quantum systems including spin systems. We choose the eigenstates of a given observable as a complete set of basis.
The generalized WvN (GWvN) entropy for a quantum state is defined with the probability distribution of this state over the chosen basis.
We show analytically that the GWvN entropy of almost all  quantum states in a Hilbert space of dimension $N$ lies around $\ln N$
with a variance of order $1/N$. This means that when $N$ is large there is typicality for  the GWvN entropy.
The GWvN entropy can also be readily defined for a mixed state, for which it has a similar typical behavior.
When it applies to a system entangled with environment, the GWvN entropy is consistent with the familiar von 
Neumann entropy.  When  the quantum state can sample adequately the Hilbert space during its dynamical evolution,
the GWvN entropy will typically change quickly from its initial value and saturate around $\ln N$, reminiscent of
the quantum H theorem~\cite{Neumann1929,Neumann2010,Han2015}.
In the sense of typicality,  the GWvN entropy is additive.  Therefore,  when it is applied a microcanonical
ensemble of $N$ quantum states in an energy shell,  the GWvN entropy is not only identical to the Boltzmann entropy
but also share its property of being additive.  In the end,  we show that the Gibbs ensemble
can be derived from  the GWvN entropy with the maximal entropy principle \cite{Jaynes1957,Neumann1927}.

\section{Generalized Wigner-von Neumann Entropy and its typicality}

In 1929, von Neumann introduced an entropy for pure quantum state but he
acknowledged that the idea was from Wigner's unpublished work~\cite{Neumann1929,Neumann2010}.
To define the entropy,  Wigner and von Neumann chose  a pair of
macroscopic position and macroscopic momentum operators that commute. Their common eigenstates
are wave functions localized on individual Planck cells in the phase space and form a complete set of basis.  A wave function
is then be mapped unitarily to the phase space with this basis.
The resulted probability distribution in the phase space is  coarse grained and used to define the WvN entropy.
This WvN entropy was generalized in Ref.\cite{Han2015} with two improvements: (1) there is no more coarse graining; (2) an
efficient way is found to compute the basis as a set of Wannier functions\cite{Fang2018}.

Here we choose an operator $A$ whose eigenstates $\left\{ \left|\phi_{i}\right\rangle \right\}$
form a complete basis. For a quantum state $\ket{\varphi}$, we use its probability distribution over
the basis $\left\{ \left|\phi_{i}\right\rangle \right\}$ to define an entropy
\begin{equation}
S(\varphi)\equiv-\sum_{j=1}^N\left|\left\langle \varphi|\phi_{j}\right\rangle \right|^{2}\ln\left|\left\langle \varphi|\phi_{j}\right\rangle \right|^{2}\,. 
\label{gwvn}
\end{equation}
Entropy of this form  has  appeared in many different contexts and had different names~\cite{Mirbach1998,Polkovnikov2011a,Santos2011,Han2015,Anza2017,Safranek2018,Wooters2018}.
In particular, without the minus sign,  it was defined as the information of operator $A$ in Ref.~\cite{Everett}. 
To our judgement,
it is fair to regard Eq. (\ref{gwvn}) as the generalization of the quantum entropy proposed by Wigner
and von Neumann in 1929~\cite{Neumann1929,Neumann2010}. We are to show in this work
that this GWvN entropy is independent of choice of $A$ in the sense of typicality.

\begin{figure}
\includegraphics{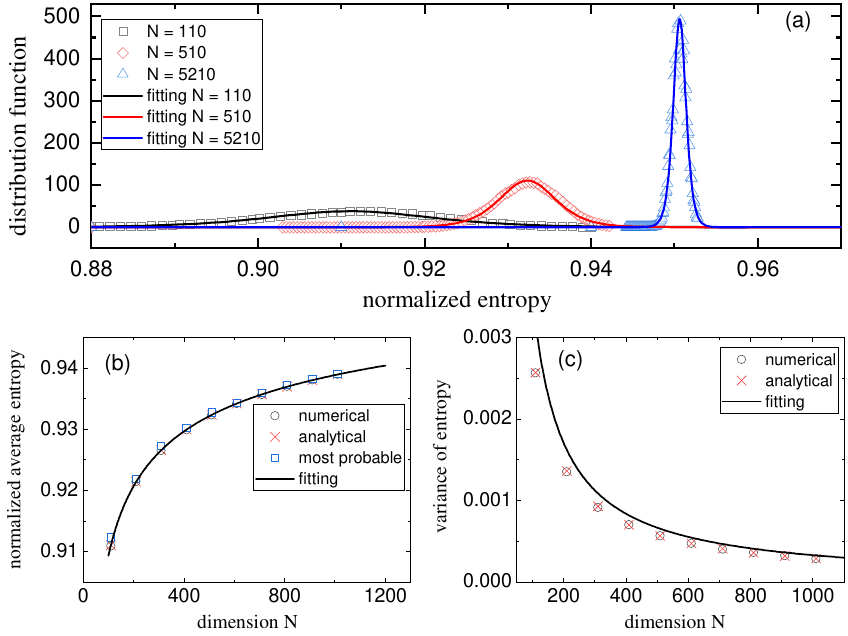}%
\caption{(color online) Statistical properties of the GWvN entropy $S$ over a Hilbert space of
dimension $N$. (a) The distributions of the normalized entropy $s=S/\ln N$. (b)
The average entropy obtained by numerical calculations, analytical results and  fittings.
The blue rectangle represents the most probable entropy. (c) The variance of the entropy obtained by numerical calculations,
analytical results and  fittings.
\label{fig1:fitting}}
\end{figure}

Due to the normalization rule, all quantum states in a Hilbert space of $N$ dimensions lie
on the $\left(2N-1\right)$-dimensional hypersphere $\mathbb{S}^{2N-1}$,
\begin{equation}
\sum_{j=1}^N|z_j|^2=1~,~~~~z_j=\braket{\phi_j|\varphi}
\end{equation}
We are interested in the statistical average, variance, and distribution of the GWvN entropy when
the quantum states are sampled on this hypersphere uniformly at random.
Let us look at the numerical results first.  Shown in Fig. \ref{fig1:fitting}(a)
are the distributions  of the entropies for $N=110,510,5210$.  The figure shows that
the distribution $P(S)$ becomes narrower and the average  gets closer to $\ln N$ quickly as
$N$ increases. These two trends are further demonstrated in Fig. \ref{fig1:fitting}(b,c).
These results indicate  that when $N$ is large, which is the usual case
for a quantum many-body system, the GWvN entropy
is the same for overwhelming majority of the quantum states with a value that is very close to $\ln N$.
It is clear that these results are  independent of the choice of operator $A$.

We find that the distribution $P(S)$ of the GWvN entropies obtained numerically in Fig. \ref{fig1:fitting}
can be fit very well with  a  function ${\cal P}(s)=d\Pr(s)/ds$,  where $\Pr(s)$ is
a Fermi-Dirac-like  function
\begin{equation}
\Pr(s)=\frac{1}{1+\exp\left[-c_{N}\left(s-\frac{\mu_{N}}{\ln N}\right)\right]} \label{eq:asydist}
\end{equation}
with $s=S/\ln N$ being the normalized GWvN entropy and the parameters
\begin{equation}
\mu_{N}=\frac{\Gamma'(N+1)}{\Gamma(N+1)}-\frac{\Gamma'(2)}{\Gamma(2)},\,~~~~c_{N}=\mu_{N}\sqrt{\frac{\uppi^{2}N}{\uppi^{2}-9}}.
\end{equation}
Here $\Gamma'(z)$ is the derivative of gamma function with respect to $z$.  As shown in Fig. \ref{fig1:fitting},
all the three essential features of the entropy distribution, shape, average value, and variance, are captured very well
by ${\cal P}(s)$. The empirical distribution in Eq.(\ref{eq:asydist}) is discussed and compared with rigorous results of 
concentration of measure and Levy's lemma in Appendix A. 

We turn to analytical results, which can be obtained for  the average value and the variance of the GWvN entropy
over all quantum states on $\mathbb{S}^{2N-1}$. This is achieved with the introduction
of an auxiliary function $H_{\lambda}=-\sum_{j}\left|z_{j}\right|^{2\lambda}$~\cite{Jones1990},
which is related to the GWvN entropy as
\begin{equation}
S(\varphi)=\left.\frac{dH_{\lambda}}{d\lambda}\right|_{\lambda=1}.
\end{equation}
We average $\left|z_{j}\right|^{2\lambda}$ over all points on $\mathbb{S}^{2N-1}$ and obtain (see Appendix B for details)
\begin{equation}
\left\langle\left|z_{j}\right|^{2\lambda}\right\rangle=\frac{\Gamma\left(N\right)\Gamma\left(1+\lambda\right)}{\Gamma\left(N+\lambda\right)}\label{eq:ppow}\,.
\end{equation}
Therefore, the average GWvN entropy reads
\begin{equation}
\langle S \rangle=\left.-N\frac{d\langle\left|z_{i}\right|^{2\lambda}\rangle}{d\lambda}\right|_{\lambda=1}=\mu_N\,.\label{eq:sav}
\end{equation}
Asymptotically, we have $\mu_N=\ln N+\gamma-1+O(1/N)$,  where   $\gamma\approx 0.577$ is the Euler-Mascheroni constant.
This confirms
our numerical and fitting results (see Fig. \ref{fig1:fitting}(b)). In fact, when $N$ is of order $10^{19}$ (about 63 spins in spin-1/2 model, still a small system), $\langle S \rangle$ is already $0.99\ln N$, less than 1\% off the maximal value.

Similarly, with  the auxiliary function $H_\lambda$, we find the variance of the GWvN  entropy according to (\ref{eq:sav})(\ref{ieqj})(\ref{ineqj})
\begin{align}
\sigma^{2}_{S}&=\left.\frac{\partial^{2}\left\langle H_{\lambda}H_{\zeta}\right\rangle}{\partial\lambda\partial\zeta}\right|_{\lambda=\zeta=1}-\langle S \rangle^{2}\notag\\&=N(N-1)\left.\frac{\partial^{2}\langle \left|z_{i}\right|^{2\lambda}\left|z_{j}\right|^{2\zeta}\rangle}{\partial\lambda\partial\zeta}\right|_{\lambda=\zeta=1,i\neq j}\notag\\&+N\left.\frac{\partial^{2}\langle \left|z_{i}\right|^{2(\lambda+\zeta)}\rangle}{\partial\lambda\partial\zeta}\right|_{\lambda=\zeta=1}-\langle S \rangle^{2}
\notag\\&=\frac{1}{N+1}\cdot\left(\frac{\pi^{2}}{3}-2\right)-\Psi_{1}\left(N+1\right),
\label{sigma}
\end{align}
where $\Psi_{1}\left(z\right)=d^{2}\ln\Gamma\left(z\right)/d z^{2}$ is trigamma function.
Asymptotically, $\Psi_{1}\left(N+1\right) =1/(N+1)+O(1/N^2)$.
So, the leading term of the  variance of GWvN entropy is $\sim 0.3/N$ \cite{Mirbach1998}.
It agrees with the numerical result and the fitting result(see Fig. \ref{fig1:fitting}(c)).

All the above results, analytical or numerical, demonstrate clearly that  there is typicality
for the GWvN entropy when the dimension of the occupied Hilbert space $N$ is large enough.
In other words, for a closed many-body quantum system,
the GWvN entropy of almost any of its quantum state is very close to
$\ln N$ and the exception is very rare. These results are independent
of choice of  operator $A$ as long as the eigenstates of $A$ forms a complete basis.
When $N$ can be regarded as the number of
microstates (eg. energy eigenstates) of a quantum system,
GWvN entropy is consistent with the well known Boltzmann entropy.
We will discuss more the physical implication of these results.

\section{GWvN entropy of subsystems}
The GWvN entropy can be extended to quantum subsystems which are described by mixed states.
We consider a closed quantum system that consists of two subsystems.
We focus on one of the subsystems and call the other subsystem environment.
The whole  system is in a pure quantum
state $\left|\varphi\right\rangle =\sum_{i,\alpha}c_{i\alpha}\left|\phi_{i},\psi_{\alpha}\right\rangle $
with normalization $\sum_{i,\alpha}\left|c_{i\alpha}\right|^{2}=1$.
Here the Roman indices represent the subsystem and the Greek indices the environment.
By convention, $\{\phi_i, i=1,2,\cdots,n\}$ is a complete basis for the subsystem
and $\{\psi_\alpha,\alpha=1,2,\cdots,m\}$ is a complete
basis for the environment.
Tracing out the environment in the density matrix of the system,
$\rho=\sum_{ij,\alpha\beta}c_{i\alpha}c_{j\beta}^{*}\left|\phi_{i},\psi_{\alpha}\right\rangle \left\langle \phi_{j},\psi_{\beta}\right|$
, we obtain the reduced density matrix for the subsystem
\begin{equation}
\rho_{s}=\text{Tr}_{e}\rho=\sum_{ij}p_{ij}\left|\phi_{i}\right\rangle \left\langle \phi_{j}\right|,
\end{equation}
where $p_{ij}=\sum_{\alpha=1}^{m}c_{i\alpha}c_{j\alpha}^{*}$.
The definition of GWvN entropy for the subsystem is
\begin{eqnarray}
S(\rho_s)&=&-\sum_{i=1}^n \text{Tr}(\rho_s\ket{\phi_{i}}\bra{\phi_{i}})\ln \text{Tr}(\rho_s\ket{\phi_{i}}\bra{\phi_{i}})\nonumber\\
&=&-\sum_{i=1}^n p_{ii}\ln p_{ii}\,.
\end{eqnarray}
We are interested in the average value and variance of $S(\rho_s)$ when the quantum state $\left|\varphi\right\rangle$
of the whole system is sampled uniformly over the hypersphere $\mathbb{S}^{2N-1}$. Here $N=nm$. 

We introduce another  auxiliary function  $K_{\lambda}=-\sum_{i=1}^{n}p_{ii}^{\lambda}$, 
which is related to the entropy as
\be
S(\rho_s)=\left.\frac{dK_{\lambda}}{d\lambda}\right|_{\lambda=1}.
\label{kf}
\ee
Direct computation (see the Appendix B for details) shows that
\begin{equation}
\left\langle p_{ii}^{\lambda}\right\rangle=\frac{\Gamma\left(mn\right)\Gamma\left(m+\lambda\right)}{\Gamma\left(mn+\lambda\right)\Gamma\left(m\right)}.\label{parsum}
\end{equation}
When $\lambda=1$, we have $\langle p_{ii}\rangle=1/n$ as expected. This leads to
\begin{equation}
\left.-\frac{d\langle p_{ii}^{\lambda}\rangle}{d\lambda}\right|_{\lambda=1}=\frac{1}{n}\left[\frac{\Gamma'(mn+1)}{\Gamma(mn+1)}-\frac{\Gamma'(m+1)}{\Gamma(m+1)}\right].
\end{equation}
With Eq.(\ref{kf}) we have the average of the GWvN entropy
\begin{equation}
\left\langle S\left(\rho_{s}\right)\right\rangle=\frac{\Gamma'(mn+1)}{\Gamma(mn+1)}-
\frac{\Gamma'(m+1)}{\Gamma(m+1)}\approx \ln n,\label{subavs}
\end{equation}
where the approximation is asymptotic and it holds when both $n$ and $m$ are very large.
The variance of $S(\rho_{s})$ can also be derived analytically. With Eqs. (\ref{subavs},\ref{ieqj},\ref{ineqj}), we have
\begin{align}
\sigma^{2}_{\rho_{s}}&=\left.\frac{\partial^{2}\left\langle K_{\lambda}K_{\zeta}\right\rangle}{\partial\lambda\partial\zeta}\right|_{\lambda=\zeta=1}-\langle S\left(\rho_{s}\right)\rangle^{2}\notag\\&=n(n-1)\left.\frac{\partial^{2}\langle p_{ii}^{\lambda}p_{jj}^{\zeta}\rangle}{\partial\lambda\partial\zeta}\right|_{\lambda=\zeta=1,i\neq j}\notag\\&+n\left.\frac{\partial^{2}\langle p_{ii}^{\lambda+\zeta}\rangle}{\partial\lambda\partial\zeta}\right|_{\lambda=\zeta=1}-\langle S\left(\rho_{s}\right)\rangle^{2}
\notag\\&=\frac{m+1}{N+1}\Psi_{1}\left(m+1\right)-\Psi_{1}\left(N+1\right),
\label{subsigma}
\end{align}
When $N> m\gg 1$, we have
\begin{equation}
\sigma^{2}_{\rho_{s}}\approx \frac{1}{2mN}+O(\frac{1}{N^2})\,.
\end{equation}
Here we have used the condition $m>n$, which is usually the case. 
This shows that the variance of a subsystem's entropy is effectively controlled by the environment
and the whole system.  The reason is that we are averaging over the Hilbert space of
the whole system, where the overwhelming majority of the quantum states are almost 
maximally entangled~\cite{Page}. If  the subsystem and the environment are not entangled, 
then the subsystem can be regarded as
an isolated system, which is already discussed in the last section. It is clear from these results 
that the GWvN entropy of a subsystem has typicality. 

Similarly, the GWvN entropy of  the environment is typical with the average being
$\left\langle S\left(\rho_{e}\right)\right\rangle\approx \ln m$ and the variance of order $1/(nN)$.

It is interesting to compare these results on mixed states with the familiar 
von Neumann entropy. For this purpose, we assume that $m\gg n$ since the environment should usually 
be much larger than the subsystem.  Page computed the average value of the von Neumann entropy 
$S_v(\rho_s)=-\text{Tr}\rho_s\ln\rho_s$ when the quantum state $\ket{\varphi}$ is sampled randomly~\cite{Page}. 
He found that the average value is approximately $\ln n-n/(2m)$, which is consistent with 
the GWvN entropy asymptotically.  Page did not compute the variance of $S_v(\rho_s)$.
This shows that the GWvN entropy for a system with a large environment 
is consistent with  the von Neumann entropy, which was 
already noticed in Ref.~\cite{Han2015}. However, for the environment, 
its von Neumann entropy $S_v(\rho_e)=S_v(\rho_s)\approx \ln n$, very different
from its GWvN entropy $\ln m$.  

Strictly, the GWvN entropy  is not additive. However, in the sense of typicality, it is additive.
This is evident in the following result
\ba
&&\left\langle S\left(\rho_{s}\right)+S\left(\rho_{e}\right)-S\left(\rho\right)\right\rangle\nonumber\\
&=&\frac{\Gamma'(mn+1)}{\Gamma(mn+1)}-\frac{\Gamma'(m+1)}{\Gamma(m+1)}-\frac{\Gamma'(n+1)}{\Gamma(n+1)}+\frac{\Gamma'(2)}{\Gamma(2)}\nonumber\\
&=& 1-\gamma+O(\frac{m+n}{mn})\,,
\label{eq:addi}
\ea
When $N$, $m$, $n$ are large, we can safely ignore the constant $1-\gamma$ and therefore have
$\braket{S(\rho_{s})}+\braket{S(\rho_{e})}=\braket{S(\rho)}$. The result can be easily generalized to multipartite systems where
each subsystem has a Hilbert space of large dimension.

The dimension $N$ of the Hilbert space in the above discussion should be regarded as
the dimension of a sub-space that is physically relevant. As an example and also as an important application,
let us consider quantum microcanonical ensemble, which is characterized by an energy shell $[E, E+\Delta E]$ with large but finite number of
energy eigenstates ~\cite{Landau1980,Huang}.
Suppose that $N$ is the number of  energy eigenstates in the shell. According to the above results,
the  GWvN entropy of any quantum state (pure or mixed) in the microcanonical ensemble is typically $\ln N$
with a very small variance.  This is consistent with the well known Boltzmann entropy.

In the above discussion, we have fixed  the basis $\left\{ \left|\phi_{i}\right\rangle \right\}$ (or operator $A$)
while sampling  the quantum states in the Hilbert space. It is clear that we can equivalently
fix the quantum state while sampling all possible bases. In fact, this is exactly what von Neumann did
in his 1929 paper\cite{Neumann1929,Neumann2010}.

\section{Dynamic evolution of GWvN entropy}
Our results for the GWvN entropy so far are kinematic and have nothing to do with the Hamiltonian
of a quantum system. In this section, we investigate how the GWvN entropy evolves dynamically and
find that the GWvN entropy has  dynamical typicality,  similar to the dynamical behavior of   observables found
in Ref. ~\cite{Bartsch2009,Reimann2018}.  For  an isolated  quantum system with a set of energy eigenstates $\ket{E_j}$,
its dynamical evolution is given by
\be
\ket{\varphi(t)}=\sum_j a_j e^{-iE_jt/\hbar}\ket{E_j}\,,
\ee
where $a_j=\braket{E_j|\Psi(0)}$ is determined by the initial quantum state $\ket{\varphi(0)}$.
In general, because $a_j$ approaches quickly to zero as $j\rightarrow\infty$, there are only a finite number of
energy eigenstates  occupied.  During the dynamical evolution, this number of occupied states does not change
as $a_j$ is independent of time.  This means that when a quantum system evolves dynamically,
it dynamical path in the Hilbert space will lie entirely in this  sub-Hilbert space of  occupied states.

Let us consider a quantum system with a macroscopic number of particles. In this case,
 the dimension $N$ of the sub-Hilbert space of occupied states is in general enormously large.
 It is natural to expect that almost all the quantum states on the dynamical path are typical and
their GWvN entropies are very close to $\ln N$ with small fluctuations.  This is exactly what is
implied in the quantum H theorems proved in Ref.\cite{Neumann1929,Neumann2010,Han2015}.
However, according to these proofs, there are exceptions that occur when
the system's Hamiltonian has a great deal of degeneracy in its eigen-energies and eigen-energy differences.
These degeneracies are shown to be closely connected to the integrability of the Hamiltonians~\cite{qmixing}.
In other words, when the system is integrable,  its quantum dynamics will be restricted by various good
quantum numbers and can not adequately sample the sub-Hilbert space. When the system is non-integrable,
its quantum dynamics can adequately sample the sub-Hilbert space so that the quantum states involved in
the dynamics are typical.

We now show how the GWvN entropy relaxes dynamically in a quantum chaotic system.
We choose a complete orthonormal basis, $\left\{ \left|\psi_0\right\rangle ,\left|\xi_{i}\right\rangle \right\}(i=1,\cdots,N-1)$,
where $\left|\psi_0\right\rangle $ is the  initial state. In this case, the GWvN entropy is zero initially.
As the system evolves  under unitary operator $U\left(t\right)=\exp\left(-iHt/\hbar\right)$, its GWvN entropy
changes with time as
\begin{equation}
 \overline{S(t)}=-\frac{d}{d \beta}\left[p^{\beta}\left(t\right)+\sum_{i=1}^{N-1}
 \overline{ \left|\langle\psi\left(t\right)\left|\xi_{i}\right\rangle \right|^{2\beta}}\right]_{\beta=1},
\end{equation}
where $p\left(t\right)=\left|\langle\psi\left|\psi\left(t\right)\right\rangle \right|^{2}$ is the surviving probability
(also called fidelity)~\cite{Flambaum2001a,Torres-Herrera2014} and the overline denotes  averaging
over the other $(N-1)$ basis $\left\{ \left|\xi_{i}\right\rangle \right\}$.  The averaging is justified by that
the system is quantum chaotic and its dynamics can sample adequately in the subspace spanned by $\left\{ \left|\xi_{i}\right\rangle \right\}$.

Note that $\left|\langle\psi\left(t\right)\left|\xi_{i}\right\rangle \right|^{2}=\left(1-p\left(t\right)\right)\cdot\left|\langle\psi'\left(t\right)\left|\xi_{i}\right\rangle \right|^{2}$, where $\left|\psi'\left(t\right)\right\rangle$ is the normalized state projected by $\left|\psi\left(t\right)\right\rangle $ onto the subspace spanned by $\left\{ \left|\xi_{i}\right\rangle \right\} $.  Similar to  Eq. \eqref{eq:ppow}, we have
\begin{equation}
\overline{ \left|\langle\psi\left(t\right)\left|\xi_{i}\right\rangle \right|^{2\beta}} =\left(1-p\left(t\right)\right)^{\beta}\frac{\Gamma\left(N-1\right)\Gamma\left(1+\beta\right)}{\Gamma\left(N-1+\beta\right)}.
\end{equation}
Eventually we find
\begin{equation}
\overline{S(t)} =f\left(p\left(t\right)\right)+\left(1-p\left(t\right)\right)\left[\frac{\Gamma'(N)}{\Gamma(N)}-\frac{\Gamma'(2)}{\Gamma(2)}\right],\label{eq:st}
\end{equation}
where $f(p)=-p\log p-(1-p)\log(1-p)$.
The deviation is negligible with the order $O(1/N)$
by typicality arguments. More precisely, it is a direct result from \eqref{sigma}. A very interesting fact is
that this expression of $S\left(t\right)$ explicitly verifies the validity of the conjectured form proposed by
Flambaum and Izrailev \cite{Flambaum2001}.

\section{connection to thermodynamics}
The GWvN entropy can also be used to derive the Gibbs ensemble with
the maximal entropy principle \cite{Jaynes1957,Neumann1927}.  We choose the operator $A=H$
and its eigenstates $\ket{E_i}$ form a complete basis.
For a typical quantum state $\rho$, its GWvN entropy is
$S=-\sum_{i}\text{Tr}\left(\rho P_{E_{i}}\right)\ln\text{Tr}\left(\rho P_{E_{i}}\right)$,
where $P_{E_{i}}=\ket{E_i}\bra{E_i}$.
We want to maximize it under the condition that $\left\langle H\right\rangle =\text{Tr\ensuremath{\left(\rho H\right)}}$ is constant.
Mathematically, this can be done as
\begin{equation}
\left.\frac{d}{d\varepsilon}\sum_{i}\text{Tr}\left(\left(\rho+\varepsilon\Omega\right)P_{E_{i}}\right)\ln\left(\text{Tr}\left(\left(\rho+\varepsilon\Omega\right)P_{E_{i}}\right)\right)\right|_{\varepsilon=0}=0,
\end{equation}
with the restriction $\text{Tr}\left(\Omega H\right)=0$ and $\text{Tr}\left(\Omega\right)=0$. This leads to
\begin{equation}
\text{Tr}\left(\Omega\sum_{i}\ln\left(\text{Tr}\left(\rho P_{E_{i}}\right)\right)P_{E_{i}}\right)=0.
\end{equation}
It can be inferred that $\sum_{i}\ln\left(\text{Tr}\left(\rho P_{E_{i}}\right)\right)P_{E_{i}}$ is linear function of $H$.
Therefore the density matrix has the general form
\begin{equation}
\rho=\alpha\exp\left(-\beta H\right),\label{eq:gge}
\end{equation}
where $\alpha$ is normailzation factor and $\beta$ is inverse temperature. Explicitly we have
\begin{equation}
\alpha=\frac{1}{\text{Tr}\left(\exp\left(-\beta H\right)\right)},\,~~~~\beta=\frac{\delta S}{\delta\left\langle H\right\rangle}.\label{eq:gentem}
\end{equation}

From standard textbook in statistical physics \cite{Landau1980,Huang}, Eq.(\ref{eq:gge}) represents the well known Gibbs ensemble. And from (\ref{eq:gentem}) we have
\begin{equation}
\Delta\left\langle H\right\rangle =\frac{\delta\left\langle H\right\rangle }{\delta S}\Delta S=T\Delta S,
\end{equation}
where the temperature $T=1/\beta$. This is exactly the first law of thermodynamics.   The similar derivations can be done for general operator $A$  and its eigenstates $\ket{\phi_i}$ with the restriction of fixed $\left\langle A\right\rangle$, then we have the generalized Gibbs ensemble $\rho=\alpha\exp\left(-\lambda_{A} A\right)$, where $\alpha$ is normailzation factor and $\lambda_{A}$ is the  generalized inverse temperature.  With the Gibbs ensemble, other thermodynamical relations can also be easily derived.
Our result here is consistent with previous results 
in Ref.~\cite{Goldstein2006,Anza2017,Anza2018}.

\section{Conclusion}
We have generalized the quantum entropy proposed by Wigner and von Neumann in 1929.
Although the definition uses a specific complete basis, the generalized Wigner-von Neumann
entropy becomes typical when the dimension $N$ of the Hilbert space is large.
We have shown analytically that when we sample a quantum state uniformly at random in the Hilbert space,
the average of its GWvN entropy is asymptotic to $\ln N$ and its  variance is of order $1/N$.
As a result, when the GWvN entropy is applied to microcanonical ensemble, it is equivalent to
the Boltzmann entropy. When it is extended to a subsystem with large environment, it is consistent with the von Neumann
entropy.  In the end,  with the maximal entropy principle we have shown that
the GWvN entropy can be used to obtain  the  Gibbs ensemble from which all thermodynamic relations can be derived.

\begin{acknowledgments}
We thank stimulating discussion with Chushun Tian. This work was supported by the The National Key Research and Development Program 
of China (Grants No.~2017YFA0303302, No.~2018YFA0305602) .

Z. H. and Z. W.  contributed equally to this work.
\end{acknowledgments}

\appendix

\section{Concentration of measure and Levy's lemma}
Concentration of measure states that 
a function that depends in a Lipschitz way on many independent variables is almost constant.
As  a special form of concentration of measure, 
Levy's lemma  is commonly used in the typicality related literature \cite{Hayden2006,Singh2016,Tian2018PRL}. 
Here we compare the  distribution of the GWvN  entropy $S$ Eq.(\ref{eq:asydist}) to Levy's lemma. 
According to Levy's lemma, for entropy
$S(\varphi)$, where $\varphi$ is a random 
point drawn uniformly from hypersphere $\mathbb{S}^{2N-1}$,  the  upper bound of the 
deviations from expected value $\langle S\rangle$ is given by
\begin{equation}
\Pr\left(\left|S\left(\varphi\right)-\langle S\rangle\right|\geq\delta\right)\leq2
\exp\left(\frac{-2N\delta^{2}}{9\pi^{3}\eta^{2}}\right)\,,\label{levylemma}
\end{equation}
where $\eta=\sup\left|\nabla_{\varphi}S\right|$ is Lipschitz constant.  Straightforward computation shows that 
\begin{align}
\eta^{2}&=4\sum_{j=1}^{N}|z_j|^2\left(\ln|z_j|^2+1\right)^2\notag\\
&=4\left(1+2\sum_{j=1}^N |z_j|^2\ln |z_j|^2+\sum_{j=1}^{N}|z_j|^2\ln^{2}|z_j|^2\right)\notag\\
&\leq4\left(1-\ln N\right)^2
\end{align}
The last inequality is true for dimension $N> e^2$, and the equality holds only when those $|z_j|^2$ are all the same. 
This means that  $\Pr\big(|S-\langle S\rangle|\geq \delta\big)$ has a  sub-gaussian tail, which presents 
stronger convergence  than  a Poisson-like tail  in the distribution   of Eq.(\ref{eq:asydist}).
However, the GWvN entropy is far away from zero only in a small region 
given by $\big|S-\mu_N\big|\sim \ln N/c_N\sim N^{-1/2}$. In this region our distribution 
is very accurate as indicated in Fig. 1(a). 

Moreover, the variance of our distribution  Eq.(\ref{eq:asydist}) is consistent with Levy's lemma.
Among all the distributions that satisfy Levy's lemma, the one that has the 
the maximal variance  should be given by 
\begin{equation}
\rho(x=|S-\langle S\rangle|)=\begin{cases}
0 & x< \sqrt{\ln 2/f_N} \\
4 f_N x \text{e}^{-f_N x^2} & x\geq \sqrt{\ln 2/f_N}
\end{cases}
\end{equation}
where \(f_N=2N/9\pi^3 \eta^2\) is the factor in exponent of Eq.(\ref{levylemma}). 
It leads to the maximal variance that Levy's lemma allows
\begin{align}
\text{Var}_{\max}&=\int_{\sqrt{\ln 2/f_N}}^{+\infty} 4 f_N x^3 \exp(-f_N x^2)dx \\
&= \frac {9\pi^3(1+\ln 2)\eta^2} {2N}
\end{align}
The variance (of order $1/N$) of our distribution 
gives a tighter bound than the above variance bound (of order $\eta^2/N$, namely $\ln^2 N/N$).

\section{Averaging in Hilbert space}
In this Appendix, we give out the derivation details that are needed in the main text, in particular,
the derivations related to  Eqs.(\ref{eq:ppow},\ref{sigma},\ref{parsum},\ref{subsigma}).


Any quantum state $\ket{\varphi}$ in an $N$-dimensional Hilbert space can be expanded in an
orthonormal basis with coefficients $z_1,\dotsm,z_N$. $z_i$ is usually a complex number and
we denote the real part and imaginary part as $x_{2i-1}$ and $x_{2i}$. Due to the normalization,
the quantum state $\ket{\varphi}$ corresponds to point $\left(x_{1},x_{2},\dotsm,x_{2N}\right)$
on the $\left(2N-1\right)$-dimensional hypersphere $\mathbb{S}^{2N-1}$.

The  average of  a function in the $N$-dimensional Hilbert space is to integrate it  on the hypersphere $\mathbb{S}^{2N-1}$.  
We can connect the integration on the sphere with an integration in the ball\cite{Baker1997,Folland2001}. 
More precisely, let $B^{n}=\left\{ x\in R^{n}:\left|x\right|\leq1\right\}$ , suppose that $f:B^{n}\rightarrow R$ 
is continuous, then
\begin{eqnarray}
&&\int_{B^{n}}f(x)dx_{1}dx_{2},\cdots,dx_{n}\nonumber\\
&=&\int_{0}^{1}r^{n-1}dr\int_{S^{n-1}}f(rs)d\sigma_{n-1}(s)\,.
\end{eqnarray}
where $d\sigma_{n-1}(s)$ denotes an element on hypersphere $S^{n-1}$.  If $f(rx)=r^{\alpha}f(x)$ is a homogeneous 
function of degree $\alpha$, we have
\begin{equation}
\int_{B^{n}}f(x)d\left(x_{1},\cdots,x_{n}\right)=\frac{1}{\alpha+n}\int_{S^{n-1}}f(s)d\sigma_{n-1}(s).
\end{equation}

Consider a function 
\begin{equation}
g\left(x_{1},\cdots,x_{2N}\right)=\prod_{k=1}^{n}T_{k}^{\alpha_{k}}\,,
\end{equation}
where $N=nm$ and 
\begin{equation}
T_{k}=\sum_{i=1}^{m}\left(x_{2(k-1)m+2i-1}^{2}+x_{2(k-1)m+2i}^{2}\right)
\end{equation}
Then the average $\left\langle g\right\rangle$ of $g\left(x_{1},\cdots,x_{2N}\right)$ on sphere $\mathbb{S}^{2N-1}$ reads 
\begin{widetext}
\begin{align}
\left\langle g\right\rangle &=\frac{1}{\varOmega_{2N-1}}\int_{\mathbb{S}^{2N-1}}\prod_{k=1}^{n}T_{k}^{\alpha_{k}}d\sigma_{2N-1}(s)\notag\\
&=\frac{\Gamma(N)}{2\pi^{N}}\left(2\alpha_{1}+2\alpha_{2}+\cdots+2\alpha_{n}+2N\right)\int_{B^{2N}}\prod_{k=1}^{n}T_{k}^{\alpha_{k}}dx_{1}dx_{2}\cdots dx_{2N}\notag\\
&=\frac{\Gamma(N)\left(\alpha_{1}+\cdots+\alpha_{n}+N\right)}{\pi^{N}}\int_{B^{2N-2m}}\prod_{k=1}^{n-1}T_{k}^{\alpha_{k}}\int_{0}^{\sqrt{1-\sum_{k=1}^{n-1}T_{k}}}\rho^{2\alpha_{n}}\rho^{2m-1}d\rho\,d\sigma_{2m-1}\,dx_{1}dx_{2}\cdots dx_{2N-2m}\notag\\
&=\frac{\pi^{m}\Gamma(N)\left(\alpha_{1}+\cdots+\alpha_{n}+N\right)}{\pi^{N}\Gamma(m)\left(m+\alpha_{n}\right)}\int_{B^{2N-2m}}\prod_{k=1}^{n-1}T_{k}^{\alpha_{k}}\left(1-\sum_{k=1}^{n-1}T_{k}\right)^{m+\alpha_{n}}
dx_{1}dx_{2}\cdots dx_{2N-2m}\notag\\
&=\frac{\Gamma(N)\left(\alpha_{1}+\cdots+\alpha_{n}+N\right)}{\pi^{N-m}\Gamma(m)\left(m+\alpha_{n}\right)}\int_{0}^{1}r^{2\left(\alpha_{1}+\cdots+\alpha_{n-1}+N-m-\frac{1}{2}\right)}\left(1-r^{2}\right)^{m+\alpha_{n}}dr\cdot\int_{\mathbb{S}^{2N-2m-1}}\prod_{k=1}^{n-1}T_{k}^{\alpha_{k}}d\sigma_{2N-2m-1}\notag\\
&=\frac{\Gamma(N)\left(\alpha_{1}+\cdots+\alpha_{n}+N\right)}{\pi^{N-m}\Gamma(m)\left(m+\alpha_{n}\right)}\frac{\Gamma\left(\alpha_{1}+\cdots+\alpha_{n-1}+N-m\right)\Gamma\left(\alpha_{n}+m+1\right)}{2\Gamma\left(\alpha_{1}+\cdots+\alpha_{n}+N+1\right)}\frac{2\pi^{N-m}}{\Gamma(N-m)}\left\langle \prod_{k=1}^{n-1}T_{k}^{\alpha_{k}}\right\rangle_{2N-2m}\notag\\
&=\frac{\Gamma(N)}{\Gamma(m)\Gamma(N-m)}\frac{\Gamma\left(\alpha_{1}+\cdots+\alpha_{n-1}+N-m\right)\Gamma\left(\alpha_{n}+m\right)}{\Gamma\left(\alpha_{1}+\cdots+\alpha_{n}+N\right)}\left\langle \prod_{k=1}^{n-1}T_{k}^{\alpha_{k}}\right\rangle_{2N-2m}\notag\\
&=\frac{\Gamma(N)}{\Gamma^{n}(m)}\frac{\Gamma\left(\alpha_{n}+m\right)\Gamma\left(\alpha_{n-1}+m\right)\cdots\Gamma\left(\alpha_{1}+m\right)}{\Gamma\left(\alpha_{1}+\cdots+\alpha_{n}+N\right)}\left\langle T_{1}^{\alpha_{1}}\right\rangle_{2m}\notag\\
&=\frac{\Gamma(N)}{\Gamma^{n}(m)}\frac{\Gamma\left(\alpha_{n}+m\right)\Gamma\left(\alpha_{n-1}+m\right)\cdots\Gamma\left(\alpha_{1}+m\right)}{\Gamma\left(\alpha_{1}+\cdots+\alpha_{n}+N\right)},   
\end{align}
\end{widetext}
where $\varOmega_{2N-1}=\frac{2\pi^{N}}{\Gamma(N)}$ is the surface area of $\mathbb{S}^{2N-1}$.

It is useful to calculate some derivatives of special forms of $g\left(x_{1},\cdots,x_{2N}\right)=\prod_{k=1}^{n}T_{k}^{\alpha_{k}}$.

\begin{multline}
\frac{d}{d\alpha_{1}}\left\langle T_{1}^{\alpha_{1}}\right\rangle =\frac{\Gamma(N)\Gamma\left(\alpha_{1}+m\right)}{\Gamma(m)\Gamma\left(\alpha_{1}+N\right)}\\\cdot\left(\Psi\left(\alpha_{1}+m\right)-\Psi\left(\alpha_{1}+N\right)\right).\label{aventropy}
\end{multline}

\begin{multline}
\frac{\partial^{2}}{\partial\alpha_{1}\partial\alpha_{2}}\left\langle T_{1}^{\alpha_{1}+\alpha_{2}}\right\rangle =\frac{\Gamma(N)\Gamma\left(\alpha_{1}+\alpha_{2}+m\right)}{\Gamma(m)\Gamma\left(\alpha_{1}+\alpha_{2}+N\right)}\\
\cdot\Big(\left(\Psi\left(\alpha_{1}+\alpha_{2}+m\right)-\Psi\left(\alpha_{1}+\alpha_{2}+N\right)\right)^{2}
\\
+\left(\Psi_{1}\left(\alpha_{1}+\alpha_{2}+m\right)-\Psi_{1}\left(\alpha_{1}+\alpha_{2}+N\right)\right)\Big).\label{ieqj}
\end{multline}

\begin{multline}
\frac{\partial^{2}}{\partial\alpha_{1}\partial\alpha_{2}}\left\langle T_{1}^{\alpha_{1}}T_{2}^{\alpha_{2}}\right\rangle =\frac{\Gamma(N)\Gamma\left(\alpha_{1}+m\right)\Gamma\left(\alpha_{2}+m\right)}{\Gamma^{2}(m)\Gamma\left(\alpha_{1}+\alpha_{2}+N\right)}\\
\cdot\Big(\left(\Psi\left(\alpha_{1}+m\right)-\Psi\left(\alpha_{1}+\alpha_{2}+N\right)\right)\\
\cdot\left(\Psi\left(\alpha_{2}+m\right)-\Psi\left(\alpha_{1}+\alpha_{2}+N\right)\right)-\Psi_{1}\left(\alpha_{1}+\alpha_{2}+N\right)\Big).\label{ineqj}
\end{multline}
Here $\Psi(z)=\Gamma'(z)/\Gamma(z)$ and $\Psi_{1}(z)=d^{2}\ln\Gamma\left(z\right)/d z^{2}$ are digamma function and trigamma function, respectively.

\bibliographystyle{apsrev}
\bibliography{WvN_entropy}

\end{document}